\documentclass[pra,superscriptaddress,twocolumn]{revtex4-2}

\usepackage{enumerate}
\usepackage{mathtools}
\usepackage{amsfonts,amssymb,amsmath}
\usepackage[]{graphics,graphicx,epsfig}
\usepackage{amsthm}
\usepackage{graphicx}
\usepackage{dcolumn}
\usepackage{natbib}
\usepackage{color}
\usepackage{multirow}
\usepackage{ulem}
\usepackage{bm}
\usepackage{url}

\usepackage{xcolor}

\begin{document}


\title{Quantum Synchronizing Words: Resetting and Preparing Qutrit States}

\author{Andrzej Grudka}
\affiliation{Institute of Spintronics and Quantum Information, Faculty of Physics and Astronomy, Adam Mickiewicz University, 61-614 Pozna\'n, Poland} 

\author{Marcin Karczewski}
\affiliation{Institute of Spintronics and Quantum Information, Faculty of Physics and Astronomy, Adam Mickiewicz University, 61-614 Pozna\'n, Poland} 

\author{Pawe{\l} Kurzy{\'n}ski}
\email{pawel.kurzynski@amu.edu.pl}
\affiliation{Institute of Spintronics and Quantum Information, Faculty of Physics and Astronomy, Adam Mickiewicz University, 61-614 Pozna\'n, Poland} 

\author{J\c{e}drzej Stempin}
\affiliation{Institute of Spintronics and Quantum Information, Faculty of Physics and Astronomy, Adam Mickiewicz University, 61-614 Pozna\'n, Poland} 

\author{Jan W{\'o}jcik}
\affiliation{Institute of Spintronics and Quantum Information, Faculty of Physics and Astronomy, Adam Mickiewicz University, 61-614 Pozna\'n, Poland} 

\author{Antoni W{\'o}jcik}
\affiliation{Institute of Spintronics and Quantum Information, Faculty of Physics and Astronomy, Adam Mickiewicz University, 61-614 Pozna\'n, Poland} 

\date{\today}


\date{\today}


\begin{abstract}

Synchronizing words in classical automata theory provide a mechanism to reset any state of a deterministic automaton to a specific target state via a carefully chosen finite sequence of transition rules. In this work, we extend the concept of synchronizing words to quantum information theory. Specifically, we show that with only two quantum channels, it is possible to bring an arbitrary qutrit state close to a designated target state. Furthermore, we demonstrate that following this reset, any pure real qutrit state can be closely approximated using the same two channels. These findings establish a quantum analogue of synchronizing words, highlighting their potential applications in constructing minimal sets of universal quantum gates capable of both resetting and preparing arbitrary states.

\end{abstract}

\maketitle


\section{Introduction}

State preparation is a fundamental aspect of quantum information protocols, playing a pivotal role in ensuring the successful execution of quantum algorithms \cite{DiVincenzo}. In idealized scenarios, it is often assumed that a quantum system starts in a well-defined pure fiducial state, such as $|0\rangle$, which is subsequently manipulated through a series of unitary quantum gates to achieve a desired target state \cite{gatesBarenco, gatesKitaev, GatesBOYKIN, gatesShi}. However, in practice, quantum systems are vulnerable to environmental noise and decoherence, meaning they are typically found in a mixed state $\rho$ rather than a pure state. As a result, the process of resetting a mixed state $\rho$ to the fiducial state $|0\rangle$ becomes essential. This process involves non-unitary transformations or the introduction of ancillary systems. Prominent techniques for such resetting include heralding via measurements \cite{measurement1, measurement2}, engineered dissipation \cite{engdiss}, algorithmic cooling \cite{algcool1, algcool2}, quantum feedback cooling \cite{feedback1, feedback2}, and measurement-induced steering \cite{steering1, steering2}. Although the methods mentioned above represent the forefront of quantum information science, quantum state resetting remains a challenging task. Therefore, it is always desirable to explore alternative approaches that could improve performance in terms of success probability, as well as the number and complexity of transformations.

The problem of state resetting is not new and arises in many fields of science. A particularly compelling approach to this problem, introduced in the classical theory of finite automata, is the concept of synchronizing words \cite{syncCer,syncEpp,sync,syncPhD,syncMSc}. These words are finite sequences of transformations that, when applied to an automaton, drive it to a specific target state, regardless of its initial state. In this work, we extend the concept of synchronizing words to the quantum domain. Specifically, we focus on three-dimensional quantum systems (qutrits) and propose a method to reset an arbitrary qutrit state close to a unique pure target state. This method relies on only two quantum channels -- one unitary and one non-unitary -- and achieves accurate approximations of the target state after just three operations. Furthermore, the same two operations can be later used to approximate any arbitrary real qutrit state.

Our method offers several advantages. First, it is deterministic: after a finite number of steps, the protocol consistently produces a good approximation of the target state. Second, it is efficient, requiring only two quantum channels, which simplifies its implementation compared to other approaches that often demand a larger set of transformations or multiple ancillary systems. In fact, these two channels can be regarded as an almost universal set of quantum gates \cite{gatesBarenco, gatesKitaev, GatesBOYKIN, gatesShi}, as they enable the preparation of a dense set of states within the system's subspace. Moreover, beyond their preparation capabilities, this set of gates can also be used for resetting, making it even more powerful than a standard set of universal gates. Finally, our protocol does not require any external control -- it is sufficient to fix a desired sequence of gates and run it. Together, these features make our approach both practical and powerful for resetting and preparing qutrit states.

This work is organized as follows. In the next section, we introduce the classical concept of synchronizing words. Following this, we extend the concept to quantum systems. We then examine a specific quantum synchronizing word for a single qutrit that utilizes only two quantum channels. Additionally, we demonstrate how the same two channels can be used to generate a close approximation of an arbitrary real qutrit state. Finally, we discuss how to extend our set of two channels to a truly universal set of qutrit gates.


\section{Classical Synchronizing Words}

\begin{figure}[t]
\includegraphics[width=0.5\textwidth]{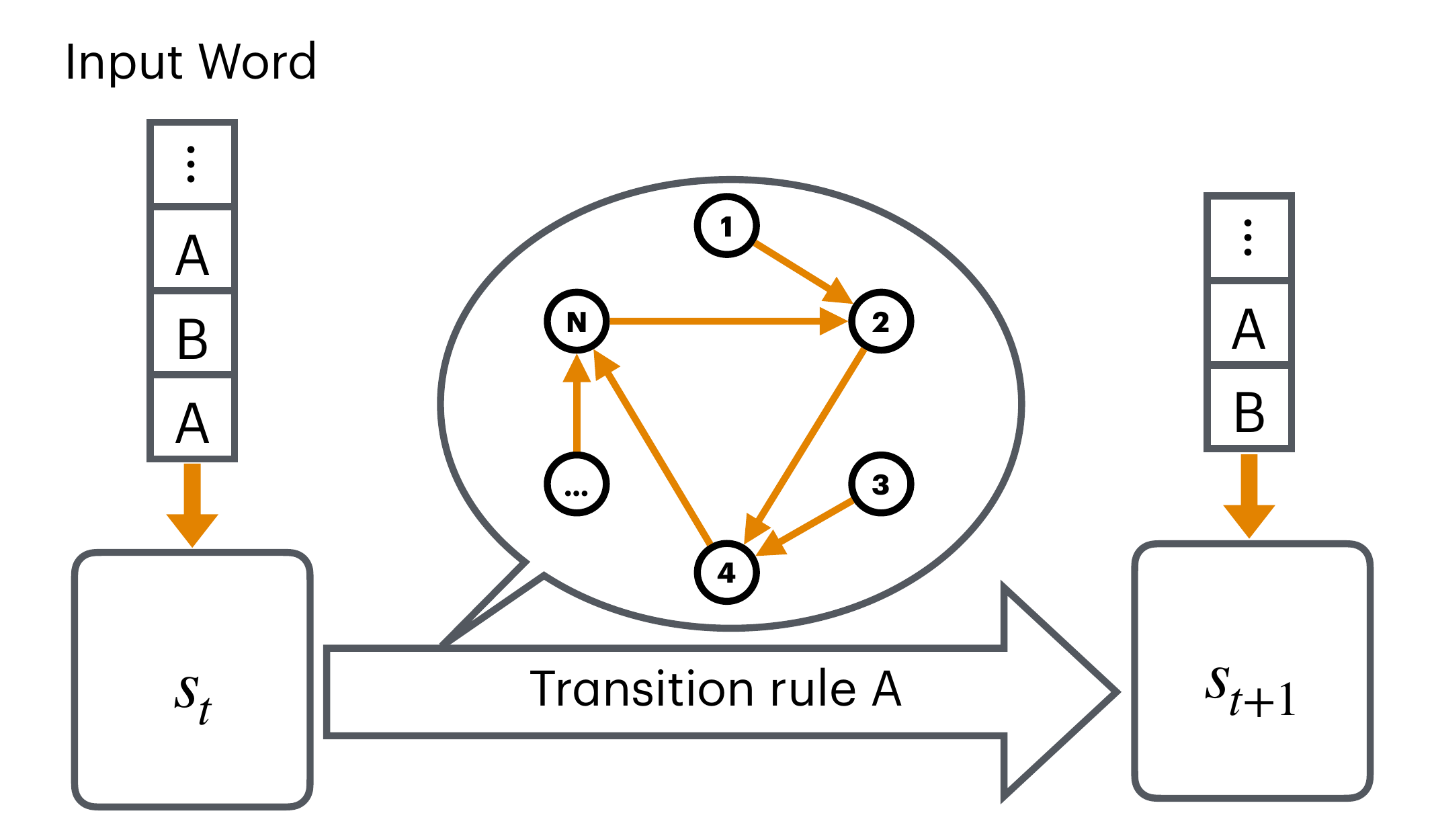}
\caption{Schematic representation of a deterministic finite automaton (DFT) and a directed graph corresponding to a transition rule.}
\label{f1}
\end{figure}

A classical deterministic finite automaton (DFA) is a system that evolves in discrete steps. It consists of $N$ possible states, denoted as $S=\{1,2,\ldots,N\}$. At any time $t$, its state is given by $s_t \in S$. Deterministic transitions between these states are governed by a finite alphabet of rules $\Sigma = \{A,B,C,\ldots,Z\}$. A DFA processes words -- sequences of rules (e.g., $ABC\ldots$) -- that drive it through a series of states:
\begin{equation}
s_0 \xrightarrow[]{A}  s_1 \xrightarrow[]{B}  s_2 \xrightarrow[]{C}  s_3 \xrightarrow[]{\dots} 
\end{equation}
Each rule is represented by a directed graph that assigns a unique descendant to each state (see Fig. \ref{f1}).   

For certain DFAs, there exist synchronizing words \cite{syncCer,syncEpp,sync,syncPhD,syncMSc}, which are specific sequences of rules that transition the automaton from any initial state to a single, specific state. Synchronizing words have numerous important applications in classical automata theory, including resetting and error correction \cite{syncPhD,syncMSc}. An interesting open problem related to synchronizing words is the \v{C}ern\'{y} conjecture \cite{syncCer}, which states that an upper bound on the length of the shortest synchronizing word for any $N$-state deterministic finite automaton is $(N-1)^2$.

\begin{figure}[t]
\vspace{-3mm}
\includegraphics[width=0.5\textwidth]{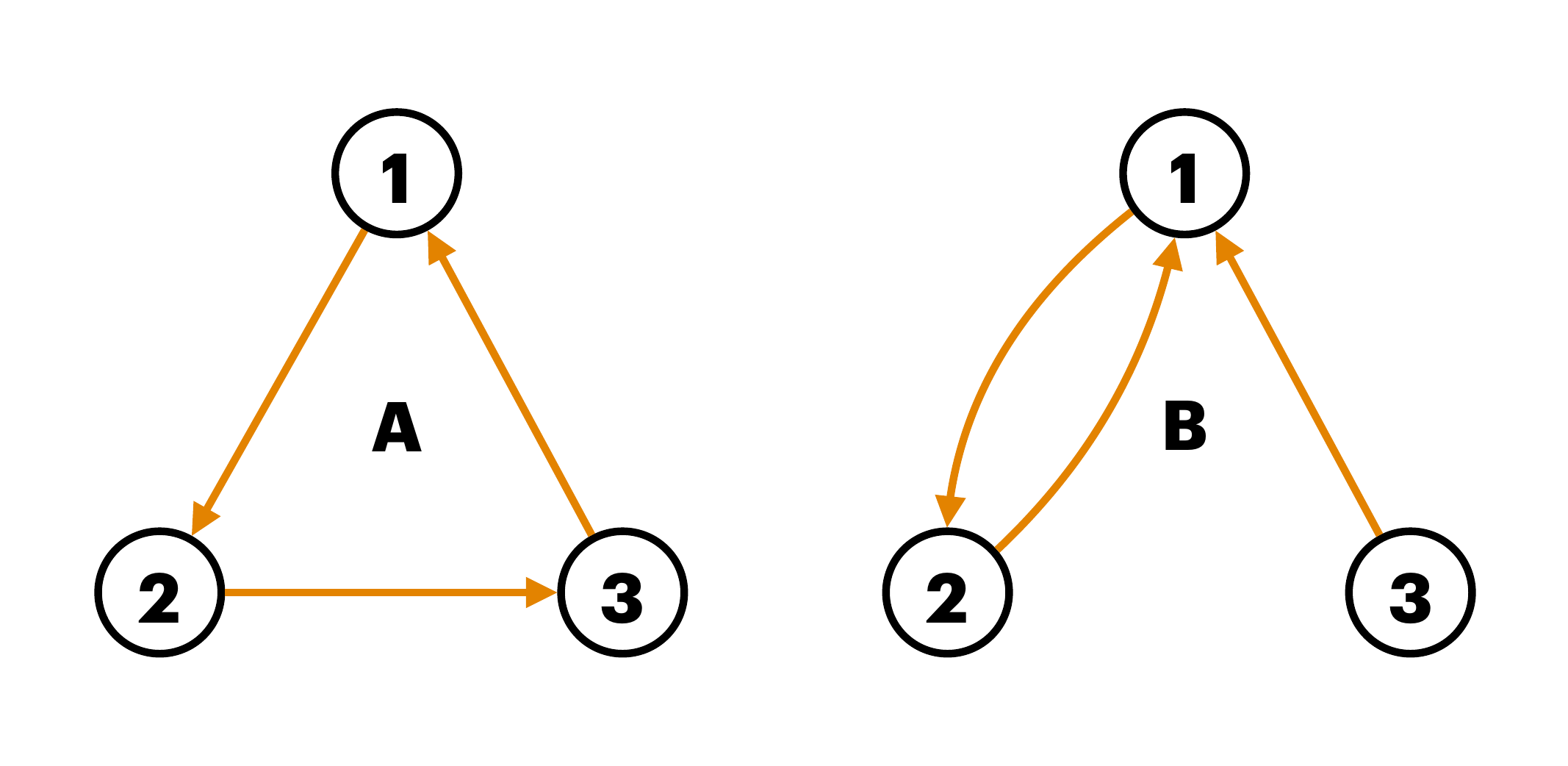}
\vspace{-10mm}
\caption{An example of a two-rule alphabet for a three-state automaton, where the word $BAB$ enables synchronization to state $1$.}
\label{f2}
\end{figure}

To gain some intuition, consider a simple DFA with three states $S=\{1,2,3\}$ and an alphabet consisting of only two rules, $\Sigma = \{A,B\}$. These rules are shown in Fig. \ref{f2}. In this case, there exists a synchronizing word, $BAB$, that transitions any state to $1$:
\begin{eqnarray}
    & & 1 \xrightarrow[]{B} 2 \xrightarrow[]{A} 3 \xrightarrow[]{B} 1, \\
     & & 2 \xrightarrow[]{B} 1 \xrightarrow[]{A} 2 \xrightarrow[]{B} 1, \\
      & & 3 \xrightarrow[]{B} 1 \xrightarrow[]{A} 2 \xrightarrow[]{B} 1.
\end{eqnarray}
We note that the existence of a synchronizing word requires at least one irreversible transition rule in the alphabet, i.e., a rule that contracts the space of states. Such rules are represented by graphs in which at least one vertex has no ancestors and at least one vertex has multiple ancestors.


\section{Quantum synchronizing words}

Let us now generalize the concept of synchronizing words to the quantum domain by introducing quantum synchronizing words. Quantum discrete finite systems (qudits) can be modeled as quantum finite automata (QFA) \cite{QFA1,QFA2}, with their states represented by a density matrix $\rho$. Notably, resetting such a system from an arbitrary state cannot be accomplished using unitary transformations alone, as unitarity preserves the von Neumann entropy of the system. To reduce the entropy of the system, at least one irreversible quantum channel is necessary.

\begin{figure}[t]
\includegraphics[width=0.4\textwidth]{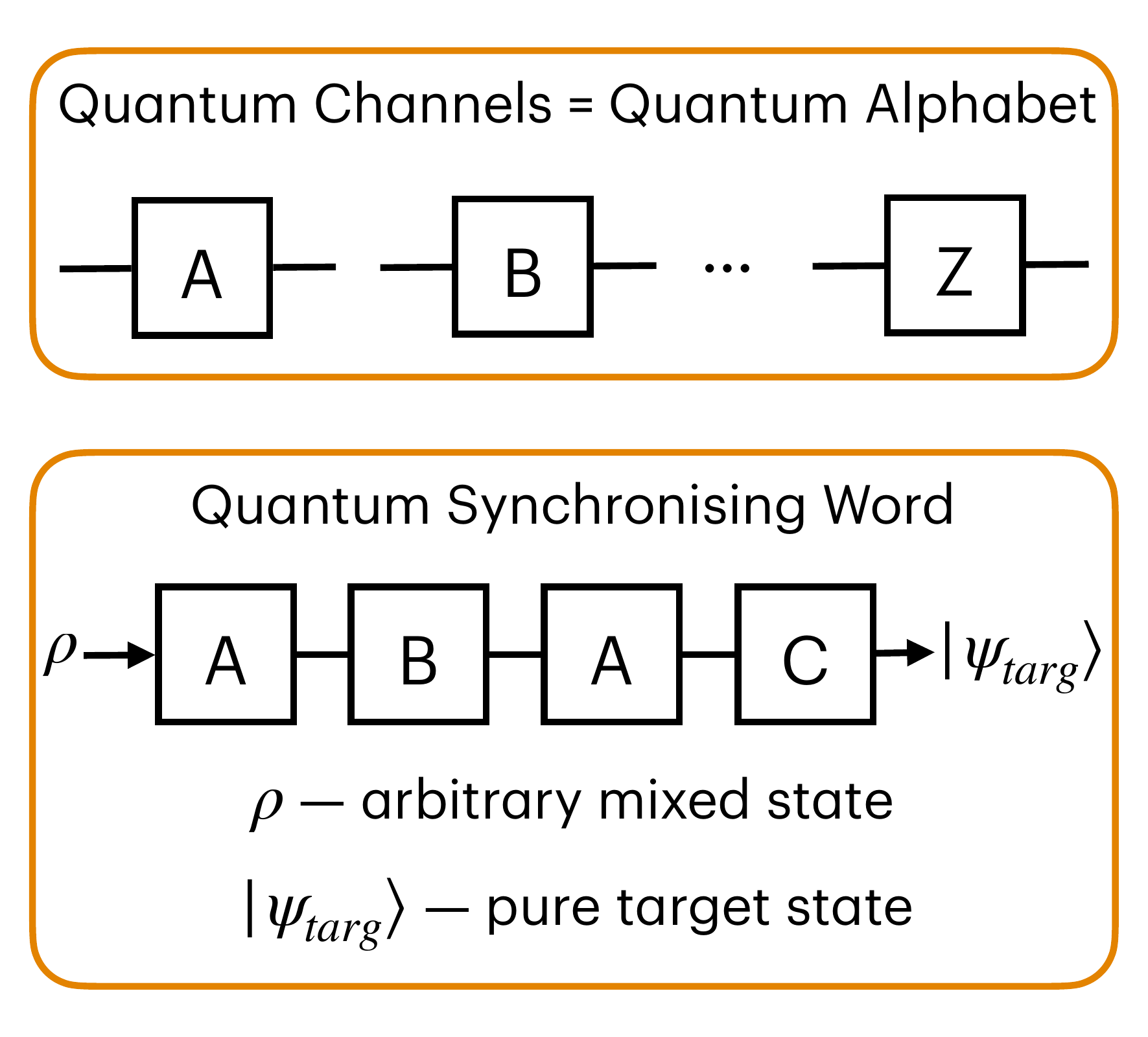}
\caption{Schematic idea of a quantum synchronizing word.}
\label{f3}
\end{figure}

Consider a finite set of quantum channels ${A, B, \ldots, Z}$ that can be applied to our system in state $\rho_0$. We refer to this set as a {\it quantum alphabet} and to its elements as {\it quantum letters}. Quantum letters can be concatenated to create finite words, such as $ABC$, whose action is given by
\begin{equation} 
ABC(\rho_0)=C(B(A(\rho_0))) = C(B(\rho_1)) = C(\rho_2) = \rho_3. 
\end{equation} 
The channels corresponding to quantum letters admit a Kraus representation \cite{Nielsen_Chuang_2010},
\begin{equation} 
A(\rho) = \sum_j A_j \rho A_j^{\dagger}, ~~~~~~\sum_j A_j^{\dagger} A_j = \openone.
\end{equation} 
We say that a quantum word is a {\it quantum synchronizing word} (QSW) if
\begin{equation} 
QSW(\rho) = |\psi_{\text{targ}}\rangle\langle \psi_{\text{targ}}|, \end{equation} 
where $\rho$ is an arbitrary state and $|\psi_{\text{targ}}\rangle$ is a particular target state. This concept is illustrated in Fig. \ref{f3}.

\section{Synchronizing word for qutrit}

Let us now focus on a qutrit and demonstrate that a two-letter alphabet $\{A, B\}$ can generate synchronizing words that drive the system to a target pure state. Our motivation for focusing on qutrits arises from the growing interest in quantum information science regarding the advantages offered by quantum systems with more than two levels. While qubits are widely regarded as the fundamental building blocks of quantum registers, qutrits represent the simplest generalization of qubits and provide unique opportunities. They are readily accessible in laboratories, appearing in systems such as photonic \cite{qutritBF} and superconducting \cite{qutritSS} platforms. Interestingly, single qutrits cannot be described classically using non-contextual models \cite{Context1, Context2, Context3, Context4}, and they offer enhanced security in cryptographic applications \cite{crypto1, crypto2}.

Consider a qutrit Hilbert space spanned by the following basis states $\{|1\rangle,|2\rangle,|3\rangle\}$. Next, consider two channels, an irreversible channel
\begin{equation}
A(\rho) = A_1 \rho A_1^{\dagger} + A_2 \rho A_2^{\dagger},
\end{equation}
where
\begin{equation}
A_1 = |2\rangle\langle 1| = \begin{pmatrix} 0 & 0 & 0 \\ 1 & 0 & 0 \\ 0 & 0 & 0 
\end{pmatrix},
\end{equation}
and
\begin{equation}
A_2 = \begin{pmatrix} 0 & 0 & 0 \\ 0 & \cos\theta & -\sin\theta \\ 0 & \sin\theta & \cos\theta 
\end{pmatrix},
\end{equation}
and a unitary channel
\begin{equation}
B(\rho) = B \rho B^{\dagger},
\end{equation}
where
\begin{equation}
B = \begin{pmatrix} \cos\varphi & -\sin\varphi & 0 \\ \sin\varphi & \cos\varphi & 0 \\ 0 & 0 & 1
\end{pmatrix}.
\end{equation}
$A_1$ projects $|1\rangle$ onto $|2\rangle$, $A_2$ generates rotation by $\theta$ in the (2,3)-plane, and $B$ generates rotation by $\varphi$ in the (1,2)-plane.

One can find that for $\theta = \varphi = \frac{\pi}{2}$, the sequence $ABA$ is a synchronizing word. More precisely,
\begin{equation}
ABA(\rho) = A(B(A(\rho))) = \rho_{ABA}= |2\rangle\langle 2|.
\end{equation}
For this choice of $\theta$ and $\varphi$, the alphabet transforms each basis state into another basis state as follows:
\begin{eqnarray}
& &A(|1\rangle\langle 1|) = |2\rangle\langle 2|,~~~~B(|1\rangle\langle 1|) = |2\rangle\langle 2|, \nonumber \\
& &A(|2\rangle\langle 2|) = |3\rangle\langle 3|,~~~~B(|2\rangle\langle 2|) = |1\rangle\langle 1|, \nonumber \\
& &A(|3\rangle\langle 3|) = |2\rangle\langle 2|,~~~~B(|3\rangle\langle 3|) = |3\rangle\langle 3|.
\end{eqnarray}
In this case, one might consider the sequence $ABA$ to be, in some sense, classical. Nevertheless, the basis states can be chosen arbitrarily. In particular, they may represent superpositions of natural states of the system that are often considered classical.

\begin{figure}[t]
\includegraphics[width=0.5\textwidth]{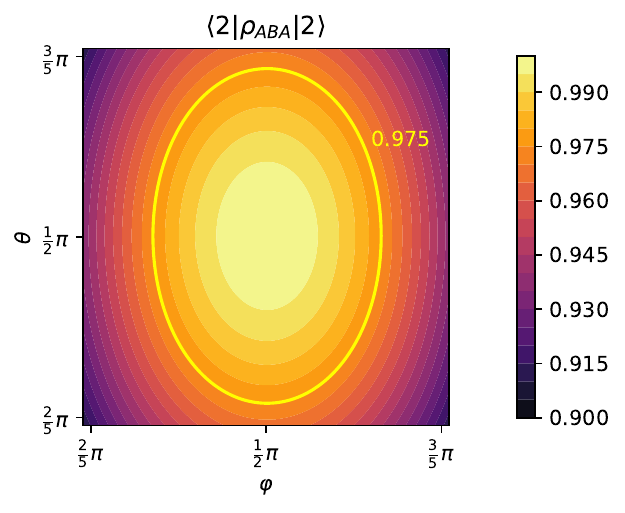}
\caption{Plot of $\langle 2|\rho_{ABA}|2\rangle$ as a function of $\theta$ and $\varphi$. The initial state was chosen to be maximally mixed.}
\label{f4}
\end{figure}

This synchronization procedure is robust with respect to small departures from the ideal angles. In the symmetric case,  $|\theta-\frac{\pi}{2}|=|\varphi-\frac{\pi}{2}|=\Delta$,  it can be easily shown that the final state satisfies 
\begin{equation}
\label{overlap}
\langle 2|\rho_{ABA}|2 \rangle\geq 1-\frac{3}{4}\Delta^2+\mathcal{O}(\Delta^4).
\end{equation}
The asymmetric choice of angles $\theta$ and $\varphi$ leads to similar results, as shown in Fig. \ref{f4}. For example, when $\theta, \varphi \in (0.45 \pi, 0.55 \pi)$ and the initial state is maximally mixed, $\rho_{ABA}$ is a good approximation of the target state, as $\langle 2|\rho_{ABA}|2 \rangle > 0.975$. Interestingly, the operators that lead to imperfect synchronization can be used in state preparation, as we will discuss in the next section.



\section{Preparation of real qutrit states}

\begin{figure}[t]
\includegraphics[width=0.5\textwidth]{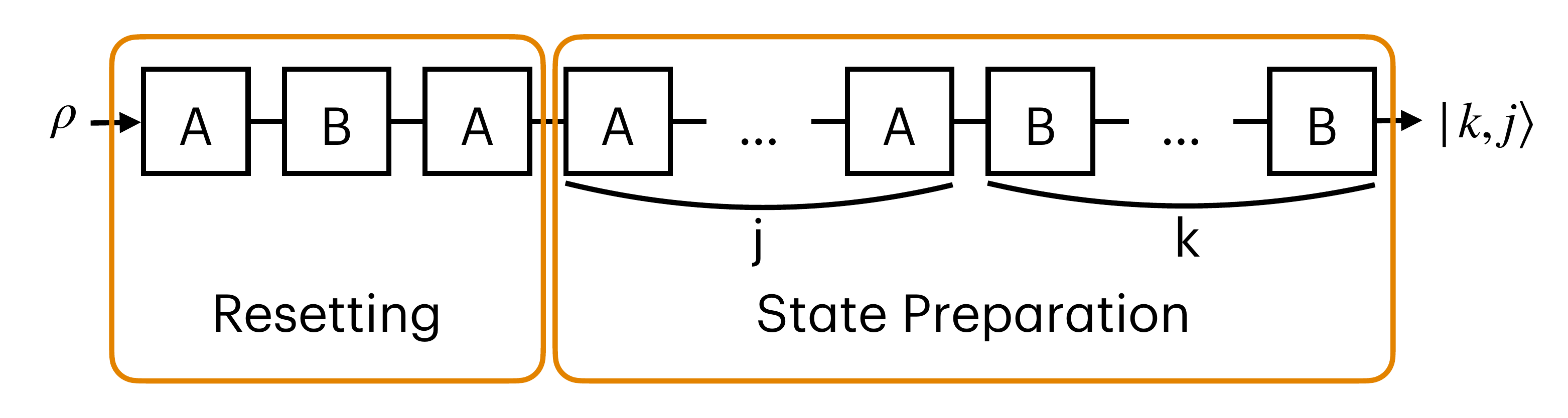}
\caption{Schematic circuit representation of a qutrit protocol for resetting and generation of states $|k,j\rangle$.}
\label{f5}
\end{figure}

\begin{figure}[htp]
\includegraphics[width=0.45\textwidth]{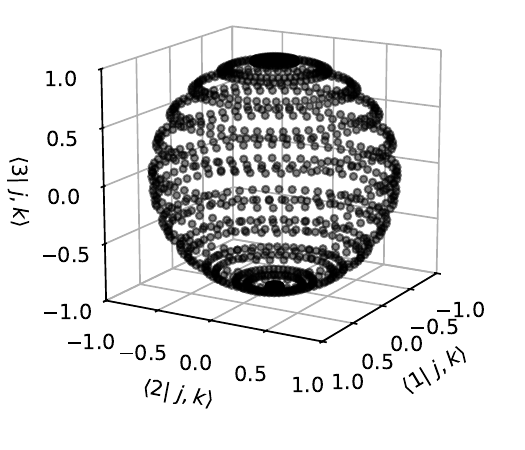}
\includegraphics[width=0.45\textwidth]{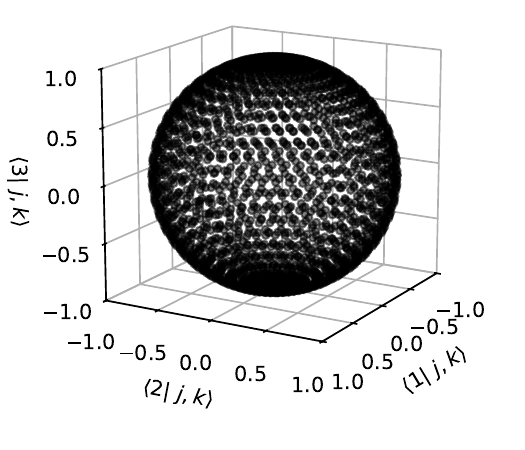}
\caption{
Set of real qutrit states $|k,j\rangle$ for: $\theta=\frac{9}{101}$, $\varphi=\frac{4}{101}\pi$ (top) and  $\theta=\frac{4}{101}$, $\varphi=\frac{4}{101}\pi$ (bottom).}
\label{f6}
\end{figure}

We now demonstrate that the previously defined alphabet, consisting of only two channels, can be employed to transform an arbitrary qutrit state into an approximation of any real pure state. Let us consider the further dynamics after reaching the state $\rho_{ABA} \approx |2\rangle\langle 2|$. The action of channel $A$ on this state is almost unitary, as
\begin{equation}
A(|2\rangle\langle 2|) = A_2 |2\rangle\langle 2| A_2^{\dagger}.
\end{equation}
Consequently, after reaching $\rho_{ABA}$, we can generate an approximation of a set of real qutrit states defined as
\begin{equation}
|k,j\rangle = B^k A_2^j |2\rangle.
\end{equation}
The schematic representation of the generating protocol is shown in Fig. \ref{f5}. The set of states ${|k,j\rangle}$ is distributed on a real unit sphere, as illustrated in Fig. \ref{f6}. Moreover, if $\theta$ and $\varphi$ are incommensurate with $\pi$, this set becomes dense.

There is an interesting trade-off between the probability of resetting, $\langle 2|\rho_{ABA}|2\rangle$, and the number of states required to achieve an approximately uniform distribution of the set ${|k,j\rangle}$ over the real sphere. As we have seen in the previous section, greater resetting precision corresponds to $|\theta-\frac{\pi}{2}|=|\varphi-\frac{\pi}{2}|=\Delta$ being closer to $0$.
However, once the state $\rho_{ABA}$ is reached, operator $A$ becomes essentially a $\theta$-rotation in the (2,3)-plane, while operator $B$ is a $\varphi$-rotation in the (1,2)-plane. Thus, the states ${B^k A_2^j|2\rangle}$ $(k,j=1,2,\ldots,n)$ will become approximately evenly distributed over the sphere for  $n=\lceil\frac{2\pi}{\Delta}\rceil$. This means that the same parameter, $\Delta^2$, is approximately proportional to the probability of failing to reset the system and to the precision of a preparation of the arbitrary pure qutrit.

\section{Discussion}

We have demonstrated that the above two-gate alphabet can be used for both resetting and preparation of real qutrit states. However, this alphabet does not constitute a fully universal set of gates, as it does not allow for the generation of complex amplitudes. To address this limitation, it can be complemented with one additional unitary element:
\begin{equation}
C = \begin{pmatrix}
1 & 0 & 0 \\
0 & e^{i\alpha} & 0 \\
0 & 0 & e^{i\beta}
\end{pmatrix},
\end{equation}
which, together with the previous two gates, can be used to approximate all qutrit states, provided that $\alpha$ and $\beta$ are incommensurable (and both are incommensurable with $2\pi$). In practical terms, for the purpose of state approximation, the strict incommensurability of $\alpha$ and $\beta$ can be replaced by the condition $a\alpha \neq b\beta$ ($a, b \in \mathbf{Z}$) for sufficiently large coprime integers $a$ and $b$. With the addition of this new gate, one can generate states of the form:
\begin{eqnarray}
& &|l,k,j\rangle = C^l|k,j\rangle = \\
& &\sin k \varphi \cos j \theta |1\rangle + e^{i l \alpha} \cos k \varphi \cos j \theta |2\rangle + e^{i l \beta} \sin j \theta |3\rangle, \nonumber
\end{eqnarray}
which approximate all pure states in the qutrit's Hilbert space.


{\it Acknowledgements.} This research is supported by the Polish National Science Centre (NCN) under the Maestro Grant no. DEC-2019/34/A/ST2/00081.

\bibliographystyle{apsrev4-2}
\bibliography{ref.bib}


\end{document}